%Paper: hep-ph/9304235
%From: My Account <me@cryptons.tamu.edu>
%Date: Thu, 8 Apr 93 10:47:27 -0800

%
% The eight figures are available upon request from me@cryptons.tamu.edu
% as one uuencoded file or separate postscript files.
%
\input harvmac

\def\Titlehh#1#2{\nopagenumbers\abstractfont\hsize=\hstitle\rightline{#1}%
\vskip .2in\centerline{\titlefont #2}\abstractfont\vskip .2in\pageno=0}

% for abstracts that won't fit on the cover page
%\def\abstractfont{\ninepoint}
%
\def\CTPa{\it Center for Theoretical Physics, Department of Physics,
      Texas A\&M University}
\def\CTPb{\it College Station, TX 77843-4242, USA}
\def\HARCa{\it Astroparticle Physics Group,
Houston Advanced Research Center (HARC)}
\def\HARCb{\it The Woodlands, TX 77381, USA}

\def\ie{\hbox{\it i.e.}}

\def\coeff#1#2{{\textstyle{#1\over #2}}}

\catcode`\@=11 % This allows us to modify PLAIN macros.

\def\lsim{\mathrel{\mathpalette\@versim<}}
\def\gsim{\mathrel{\mathpalette\@versim>}}
\def\@versim#1#2{\vcenter{\offinterlineskip
    \ialign{$\m@th#1\hfil##\hfil$\crcr#2\crcr\sim\crcr } }}
\def\boxit#1{\vbox{\hrule\hbox{\vrule\kern3pt
      \vbox{\kern3pt#1\kern3pt}\kern3pt\vrule}\hrule}}

\def\t1{{\tilde 1}}

\def\JL{J. L. Lopez}
\def\HP{H. Pois}
\def\XW{X. Wang}
\def\AZ{A. Zichichi}
\def\DVN{D. V. Nanopoulos}

\def\slp{p\hskip-5.5pt/\hskip2pt}
\def\mpt{p_T\hskip-10.75pt/\hskip5pt}

\def\GeV{\,{\rm GeV}}
\def\TeV{\,{\rm TeV}}

\def\pb{\,{\rm pb}}

\def\ipb{\,{\rm pb^{-1}}}

\def\NPB#1#2#3{Nucl. Phys. B {\bf#1} (19#2) #3}
\def\PLB#1#2#3{Phys. Lett. B {\bf#1} (19#2) #3}

\def\PRD#1#2#3{Phys. Rev. D {\bf#1} (19#2) #3}
\def\PRL#1#2#3{Phys. Rev. Lett. {\bf#1} (19#2) #3}
\def\PRT#1#2#3{Phys. Rep. {\bf#1} (19#2) #3}

\def\TAMU#1{Texas A \& M University preprint CTP-TAMU-#1}

\nref\minimal{R. Arnowitt and P. Nath, \PRL{69}{92}{725}; P. Nath and R.
Arnowitt, \PLB{287}{92}{89} and \PLB{289}{92}{368};
\JL, \DVN, and \AZ, \PLB{291}{92}{255};
\JL, \DVN, and \HP, \PRD{47}{93}{2468};
\JL, \DVN, \HP, and \AZ, \PLB{299}{93}{262}.}
\nref\LNZb{\JL, \DVN, and \AZ, \TAMU{68/92},
CERN-TH.6667/92, and CERN-PPE/92-188.}
\nref\LNWZ{\JL, \DVN, \XW, and \AZ, \TAMU{76/92}, CERN/LAA/92-023,
and CERN-PPE/92-194.}
\nref\LNPWZ{\JL, \DVN, \HP, \XW, and \AZ, \TAMU{89/92}, CERN/LAA/93-01,
CERN-TH.6773/93, and CERN-PPE/93-16.}
\nref\HERAbooks{Physics at HERA: proceedings of the Workshop on Physics
at HERA, Hamburg, Germany, Oct. 29-30, 1991. Edited by W. Buchmuller,
G. Ingelman (DESY, Hamburg, Germany, 1992).}
\nref\ALTA{G. Altarelli, G. Martinelli, B. Mele, and R. R\"uckl,
\NPB{262}{85}{204}.}
\nref\DREES{M. Drees, and D. Zeppenfeld, \PRD{39}{89}{2536}.}
\nref\JAP{H. Tsutsui, K. Nishikawa, and S. Yamada, \PLB{245}{90}{663}.}
\nref\WW{C. F. Weizs\"acker, Z. Phys. {\bf{88}}, (1934)612;
E. J. Williams, Phys. Rev. {\bf{45}}, (1934)729.}
\nref\LN{For a review see, A. B. Lahanas and D. V. Nanopoulos,
\PRT{145}{87}{1}.}
\nref\aspects{S. Kelley, \JL, \DVN, H. Pois, and K. Yuan, \TAMU{16/92} and
CERN-TH.6498/92 (to appear in Nucl. Phys. B).}
\nref\GUNION{J. F. Gunion, and H. E. Haber, \NPB{272}{86}{1}.}
\nref\MORFIN{J. G. Morfin, and W. K. Tung, Z. Phys. C {\bf{52}} (1991) 13.}

\nfig\I{The Feynman diagrams contributing to the production channels
$e^-p\to\tilde e^-_{L,R}\chi^0_{1,2}+p$ and, $e^-p\to\tilde\nu_e\chi^-_1+p$,
through the elastic processes. The relevant deep-inelatic diagrams can
be obtained by simply replacing {\it proton} by {\it parton}.}
\nfig\III{The elastic cross section for $e^-p\to\tilde e^-_R\chi^0_1
\to ep+\slp$ versus $m_{\tilde e_R}$ (top row) and $e^-p\to\tilde
e^-_{R,L}\chi^0_{2,1}\to ep+\slp$ (bottom row). Note the dominance of the
former. The corresponding cross section for $\tilde e_L\chi^0_2$ is
negligible.}
\nfig\IV{The deep-inelastic (DI) cross section for $e^-p\to\tilde
e^-_R\chi^0_1\to eX+\slp$ versus $m_{\tilde e_R}$ (top row) and $e^-p\to\tilde
e^-_{R,L}\chi^0_{2,1}\to ep+\slp$ (bottom row). Note the dominance of the
former. The corresponding cross section for $\tilde e_L\chi^0_2$ is
negligible.}
\nfig\V{The most likely value of the tranverse momentum of the daugther
electron (weighed by the various elastic cross sections) versus the total
elastic cross section for selectron-neutralino production. The daugther
electron will be hard and with large $p_T$.}
\nfig\VI{The most likely value of the relative proton energy loss in elastic
processes (weighed by the various elastic cross sections) versus the total
elastic cross section for selectron-neutralino production (top row) and
$m_{\tilde e_R}$ (bottom row). The Leading Proton Spectrometer (LPS) will
allow determination of $\tilde z_{min}$, and thus an indirect measurement
of $m_{\tilde e_R}$.}
\nfig\VII{The elastic and deep-inelastic (DI) cross sections for
$ep\to\tilde\nu_e\chi^-_1\to ep(X)+\slp$ versus $m_{\chi^\pm_1}$. Note the
faster drop off of the deep-inelatic cross section. This same phenomenon
occurs for selectron-neutralino production.}
\nfig\VIII{The elastic plus deep-inelastic selectron-neutralino cross section
versus $m_{\tilde e_R}$ (top row). This signal will be the dominant one
for a right-handedly polarized electron beam. Also (bottom row) the elastic
plus deep-inelastic total supersymmetric cross section (including
selectron-neutralino and sneutrino-chargino channels) versus $m_{\tilde e_R}$,
showing the discovery potential at HERA on this mass variable.}
\nfig\IX{The discovery potential at HERA (\ie, the total supersymmetric
cross section) for the lightest neutralino (top row) and the sneutrino (bottom
row).}

%\special{landscape}
%\baselineskip=20pt plus 2pt minus 2pt %double space for PRD
%\centerline{EUROPEAN ORGANIZATION FOR NUCLEAR RESEARCH}
%\medskip
%\Titlehhh{\vbox{\baselineskip12pt
%\hbox{CERN-PPE/93--??}
%\hbox{?? April, 1993}
\Titlehh{\vbox{\baselineskip12pt
\hbox{CERN/LAA/93--19}
\hbox{CTP--TAMU--15/93}\hbox{ACT--05/93}}}
{\vbox{\centerline{SUSY Signals at HERA in the No-Scale}
\centerline{Flipped SU(5) Supergravity Model}}}
\centerline{JORGE~L.~LOPEZ$^{(a)(b)}$, D.~V.~NANOPOULOS$^{(a)(b)}$,
XU WANG$^{(a)(b)}$, and A. ZICHICHI$^{(c)}$}
\smallskip
\centerline{$^{(a)}$\CTPa}
\centerline{\CTPb}
\centerline{$^{(b)}$\HARCa}
\centerline{\HARCb}
\centerline{$^{(c)}${\it CERN, Geneva, Switzerland}}
\vskip .2in
%\vskip .1in

\centerline{ABSTRACT}
Sparticle production and detection at HERA are studied
within the recently proposed no-scale flipped $SU(5)$ supergravity model.
Among the various reaction channels that could lead to sparticle
production at HERA, only the following are within its limit of sensitivity
in this model: $e^-p\to \tilde e^-_{L,R}\chi^0_i+X,
\tilde \nu_e\chi^-_1+X$, where $\chi^0_i(i=1,2)$ are the two lightest
neutralinos and $\chi^-_1$ is the lightest chargino. We
study the elastic and deep-inelastic contributions to the cross sections
using the Weizs\"acker-Williams approximation. We find that the most
promising supersymmetric production channel  is right-handed selectron
($\tilde e_{R}$) plus first neutralino ($\chi^0_1$), with one hard electron
and missing energy signature. The $\tilde\nu_e\chi^-_1$ channel leads to
comparable rates but also allows jet final states. A right-handedly polarized
electron beam at HERA would shut off the latter channel and allow
preferentially the former one. With an integrated luminosity of ${\cal
L}=100\ipb$, HERA can extend the present LEPI lower bounds on $m_{\tilde e_R},
m_{\tilde\nu_e},m_{\chi^0_1}$ by $\approx25\GeV$, while ${\cal L}=1000\ipb$
will make HERA competitive with LEPII. We also show that the Leading Proton
Spectrometer (LPS) at HERA is an excellent supersymmetry detector which
can provide indirect information about the sparticle masses by measuring the
leading proton longitudinal momentum distribution.
%\bigskip
\Date{April, 1993}

\newsec{Introduction}

The search for supersymmetric (SUSY) particles using existing facilities
is the crucial problem for particle physicists nowadays. One of the most
important reasons to study detailed spectra and properties of the expected
SUSY particles on the basis of well motivated theoretical concepts,
is that quite a few particle accelerators are either running (Tevatron, LEPI,
SLC, HERA) or will become operational in the near future (LEPII) and their
center-of-mass energy is within the range of the sparticle masses. Using two
well motivated supersymmetric (the minimal $SU(5)$ \minimal\ and the no-scale
flipped $SU(5)$ \LNZb\ supergravity) models,
we have previously discussed the possible SUSY production channels and
detection signatures at the Tevatron \LNWZ\ and at LEPII \LNPWZ.
In this paper we continue this program applying it to the HERA $e^-p$ collider
within the context of the same two models.
Fortunately or unfortunately, the minimal $SU(5)$ supergravity model
is out of the reach of HERA because the slepton and squark masses
($\gsim 500\GeV$) are too large to be kinematically accesible.
On the other hand, in the no-scale flipped $SU(5)$ supergravity model, the
slepton and squark masses are much lighter and part of the parameter space
can be explored at  HERA. However, since the squark masses are always above
$200 \GeV$, the much studied production channels involving squarks \HERAbooks\
are highly suppressed and are neglected in this paper. Therefore, we focus on
the production of sleptons, charginos, and neutralinos at HERA within the
predictions of the
no-scale flipped $SU(5)$ supergravity model. It is interesting to remark that
in contrast with ``generic" supersymmetric models where the squarks can
arbitrarily be taken to be light or heavy, this is not an option in this model,
that is, HERA should {\it not} produce squarks if this model is correct.

The production processes of interest at HERA are
\eqna\I
$$\eqalignno{e^-p &\to \tilde e^-_{L,R}\chi^0_{1,2}+X, &\I a\cr
             e^-p     &\to \tilde \nu_e\chi^-_1+X, &\I b\cr}$$
both of which have very small Standard Model backgrounds. Indeed,
$\sigma(ep\to\nu_e W+p,\ W\to e\bar\nu_e)\approx10^{-3}\pb$ and $\sigma(ep\to
eZ+p,\ Z\to\nu\bar\nu)\approx7\times10^{-3}\pb$ \ALTA. Moreover, by measuring
the total $\nu_e W$ and $eZ$ cross sections through the other decay modes of
the $W$ and $Z$ one could in principle subtract off these backgrounds \DREES.
The processes in Eq. \I{} receive elastic ($Q^2<4 \GeV$), deep-inelastic
($Q^2>4\GeV$), and inelastic contributions, where $-Q^2$ is the exchanged
virtual photon mass squared. It has been shown \JAP\ that the cross section for
the inelastic processes, whereby the proton gets excited into various
resonances, is smaller than that for the other two. We neglect its contribution
in our calculations. This makes our results conservative as far as the
sparticle mass lower bound that can be explored at HERA is concerned. Also, the
exact calculation of the total cross section for the processes mentioned above
usually involves the numerical evaluation of a three- (or more) body phase
space which is rather time-consuming because of the large size of the parameter
space to be scanned. For this reason we use the Weizs\"acker-Williams (WW) \WW\
approximation scheme proposed in Refs. \refs{\ALTA, \DREES}. In this method the
$e\gamma$ reaction is treated as a subprocess with a real (on-shell) photon. By
incorporating the density distribution of photons inside protons or quarks, one
can get reasonable approximations to the total cross sections.

The signature for selectron-neutralino production is dominated by
$\tilde e_R\chi^0_1$ and consists of one outgoing hard electron plus
missing transverse momentum ($\mpt$). There is a small contribution from
$\tilde e_R\chi^0_2$ production which can produce trilepton $(\chi^0_2 \to
l^+l^-+\chi^0_1)$ or mixed $(\chi^0_2 \to 2{\rm jets}+\chi^0_1)$ signals.
Chargino-sneutrino production can also lead to one outgoing hard lepton since
the chargino can decay leptonically and the sneutrino decays mostly invisibly
$(\tilde \nu_e \to \nu_e + \chi^0_1)$

This paper is organized as follows. First we discuss the features
of the no-scale flipped $SU(5)$ supergravity model (Sec. 2). Then we give the
exact formulae for the relevant tree-level cross sections in the WW
approximation (Sec. 3), followed by the results of the calculation (Sec. 4).
Finally we dicuss the phenomenological implications of our work (Sec. 5).

\newsec{The no-scale flipped $SU(5)$ supergravity model {\rm\LNZb}}
This supersymmetric model can be viewed as a specific subset of the minimal
supersymmetric standard model (MSSM), in that its three-dimensional parameter
space is contained in the 21-dimensional parameter space of the MSSM.
This subset is not arbitrary, but determined by the application of
several well motivated theoretical constraints. In this model it is assumed
that below the Planck scale the gauge group is flipped $SU(5)$, with some
special properties  expected from a superstring-derived model,
that is, it is a string-inspired model. For example, gauge coupling unification
occurs at the scale $M_U=10^{18}\GeV$, in contrast with $10^{16}\GeV$ for
the minimal $SU(5)$ model. Moreover, the usual supergravity-induced universal
soft-supersymmetry breaking parameters are assumed to obey $m_0=A=0$, as is
the case in typical no-scale supergravity models \LN. Thus the only three
parameters  in this model are: the top-quark mass ($m_t$), the ratio of
Higgs vacuum expectation values ($\tan\beta$), and the gluino mass ($m_{\tilde
g}$). Through the running of the renormalization group equations and the
minimization of the one-loop effective potential, one can obtain the whole
set of masses and couplings (including the one-loop corrected Higgs boson
masses) in this model for each allowed point in parameter space \aspects.
In what follows we take $m_t=100,130,160\GeV$, for which we find
$2<\tan\beta<32$.

Clearly, the several sparticle masses will be correlated, and are found to
scale with the gluino mass. Of great relevance is the fact that the present
body of phenomenological constraints on the sparticle masses disallows
certain combinations of the parameters, in particular one obtains
\eqn\FI{m_{\tilde g}\approx m_{\tilde q}\gsim200\GeV.}
Also, most of the weakly interacting sparticles cannot be too heavy. In fact,
we take the ``no-scale inspired" condition $m_{\tilde g,\tilde q}\lsim1\TeV$
to hold. One finds
\eqna\FII
$$\eqalignno{m_{\tilde e_R}&<190\GeV,\quad m_{\tilde e_L}<305\GeV,
\quad m_{\tilde\nu}<295\GeV,\quad m_{\tilde\tau_1}<185\GeV,
\quad m_{\tilde\tau_2}<315\GeV,\cr
m_h&<135\GeV,\quad m_{\chi^0_1}<145\GeV,\quad m_{\chi^0_2}<285\GeV,
\quad m_{\chi^\pm_1}<285\GeV.&\FII{}\cr}$$
There are also simple approximate relations that these masses obey, namely
\eqna\FIII
$$\eqalignno{m_{\tilde e_L}&\sim0.3\,m_{\tilde g};
\quad m_{\tilde e_R}\sim0.18\,m_{\tilde g},&\FIII a\cr
m_{\chi^0_1}&\sim\coeff{1}{2}m_{\chi^0_2};\quad
m_{\chi^0_2}\approx m_{\chi^\pm_1}\sim0.3m_{\tilde g}.&\FIII b\cr}$$
For low $m_{\tilde g}$, the sneutrino mass is close to $m_{\tilde e_R}$; as
$m_{\tilde g}$ grows, the sneutrino mass approaches $m_{\tilde e_L}$.
Note that $m_{\tilde e_R}/m_{\tilde e_L}\approx0.6$, in sharp contrast with
usual approximation of degenerate selectron masses. For more details on the
construction of this model we refer the reader to Ref. \LNZb.

\newsec{The allowed production processes}
The relevant Feynman diagrams for the sparticle production channels in Eq. \I{}
are shown in Fig. 1 for the elastic contributions. The deep-inelastic processes
receive contributions analogous to those shown in Fig. 1 with the replacement
{\it proton} for {\it parton}, plus additional production diagrams involving
squark exchanges. Since it has been shown \JAP\ that the squark contributions
to the cross sections for this type of deep-inelastic processes are negligible
for $m_{\tilde q}\gsim200\GeV$, which is the case in this model (see Sec. 2),
in what follows we neglect all diagrams involving squarks. We also remark that
in this model the masses of the right- and left-handed selectrons are highly
non-degenerate (see Sec. 2), in sharp contrast with the approximation of
degenerate selectron masses usually made in the literature. The squared
amplitude for the subprocess $e\gamma \to \tilde e_{L,R}\chi^0_1$
(for unpolarized incident electrons) is given by
\eqna\II
$$\eqalignno{|{\cal M}_{L,R}|^2=\coeff{1}{2}e^6 f^2_{L,R}{1\over Q^4}
\Biggl\{
&{4\over \hat s^2}(m^2_{\chi^0_i}-m^2_{\tilde e_{L,R}})p^\mu_1p^\nu_1+
\left[4{(m^2_{\chi^0_i}-m^2_{\tilde e_{L,R}})\over (\hat u -m^2_{\tilde
e_{L,R}})^2} -
{4q^2\over \hat s(\hat u-m^2_{\tilde e_{L,R}})}\right]p^\mu_2p^\nu_2\cr
&+\left[{q^2\over \hat s^2}(m^2_{\chi^0_i}-m^2_{\tilde e_{L,R}})+
{1\over \hat s}(\hat t-m^2_{\chi^0_i})\right]g^{\mu\nu}\cr
&+{{2[2(m^2_{\chi^0_i}-m^2_{\tilde e_{L,R}})+q^2]}\over {\hat s(\hat
u-m^2_{\tilde e_{L,R}}})}(p^\mu_1p^\nu_2+p^\nu_1p^\mu_2)\Biggr\}{\cdot}
\coeff{1}{2}(-g_{\mu\nu}),&\II{}\cr}$$
where $p_1\,(p_2)$ is the electron (selectron) momentum, and
$\hat s, \hat t=(p_1-p_2)^2$, $\hat u$ are the Mandelstam variables for this
subprocess. The coupling factors $f_{L,R}$ are
\eqna\III
$$\eqalignno{f_L&={\sqrt2\over e}\left[eN^\prime_{i1}+{g\over \cos{\theta_W}}
(\coeff{1}{2}-\sin^2{\theta_W})N^\prime_{i2}\right],&\III a\cr
f_R&=-{\sqrt2\over e}\left[eN^\prime_{i1}-{{g\sin^2{\theta_W}}\over
\cos{\theta_W}}N^\prime_{i2}\right],&\III b\cr}$$
with
\eqn\IV{N^\prime_{i1}=N_{i1}\cos{\theta_W}+N_{i2}\sin{\theta_W},\qquad
N^\prime_{i2}=-N_{i1}\sin{\theta_W}+N_{i2}\cos{\theta_W},}
where $N_{i1}$, $N_{i2}$ are elements of the matrix diagonalizing the
neutralino mass matrix. Here we follow the conventions of Ref. \GUNION.
The squared amplitude of the subprocess $e\gamma\to \tilde \nu_e\chi^-_1$ is
given by
\eqna\VII
$$\eqalignno{|{\cal M}|^2=\coeff{1}{2}e^4 f^{\prime 2}_L{1\over Q^4}
\Biggl\{
&\left[{4\over \hat s^2}(m^2_{\chi^-_1}-m^2_{\tilde\nu_e})+
{{4q^2}\over {\hat s(\hat u-m^2_{\chi^-_1})}}\right]p^\mu_1p^\nu_1+
\left[{{4(m^2_{\chi^-_1}-m^2_{\tilde\nu_e})}\over {(\hat u -m^2_{\chi^-_1})^2}}
+{{4q^2}\over {\hat s(\hat u-m^2_{\chi^-_1})}}\right]p^\mu_2p^\nu_2\cr
&+\left[\hat s(\hat u-m^2_{\chi^-_1})+q^2(m^2_{\chi^-_1}-
m^2_{\tilde\nu_e})\right]
\left[{1\over {\hat s}} +{1\over {\hat u-m^2_{\chi^-_1}}}
\right]^2g^{\mu\nu}\cr
&+{{4(m^2_{\chi^-_1} -m^2_{\tilde\nu_e} -q^2)}\over {\hat s(\hat
u-m^2_{\chi^-_1})}}(
p^\mu_1p^\nu_2+p^\nu_1p^\mu_2)\Biggr\}{\cdot}
\coeff{1}{2}(-g_{\mu\nu}), &\VII{}\cr}$$
where $p_1\,(p_2)$ is the electron (chargino) momentum, and
$f^\prime_{L}=gV_{11}$, with $V_{11}$ is an element of the matrix
diagonalizing the chargino mass matrix \GUNION.

The Weis\"acker-Williams (WW) approximation \WW\ is now used to simplify
the calculation. For elastic
processes we use the following photon distribution in the proton \DREES
\eqn\VIII{{f_{\gamma|p}}(z)={\alpha\over {2\pi z}}[1+(1-z)^2]
\left[{\ln}A-{11\over 6}+{3\over A}-{3\over {2A^2}}+{1\over {3A^2}}\right],}
where $A=1+(0.71 \GeV^2)/Q^2_{min}$ and
\eqn\IX{Q^2_{min}=-2m^2_p+{1\over {2s}}\left[(s+m^2_p)(s-\hat s+m^2_p)-
(s-m^2_p)\sqrt{(s-\hat s-m^2_p)^2-4m^2_p\hat s}\,\right].}
The total elastic cross section for $ep\to Xp$ can then be written as
\eqn\X{{\sigma_{elastic}(ep\to Xp)}=\int^{z_{max}}_{z_{min}}dz
f_{\gamma|p}(z)\hat \sigma(\hat s),}
where $\hat \sigma(\hat s)$ is the total subprocess cross section for the
real $\gamma e\to X$ process (\ie, Eqs. \II{}, \VII{} integrated over the
$X$-phase space), and $z=\hat s/s$, where $\hat s$ is the
center-of-mass energy of the subprocess. For a two-body final state $X$
with particles of masses $\tilde m_1$ and $\tilde m_2$, one has
\eqn\Xa{z_{min}={1\over s}(\tilde m_1+\tilde m_2)^2.}
Also, $z_{max}=(1-m_p/\sqrt{s})^2$. For the deep-inelastic processes we use the
photon distribution in the quark of Ref. \ALTA,
\eqn\XI{{P_{\gamma|q_f}}(\eta)={\alpha\over {2\pi}}e^2_{q_f}{{1+
(1-\eta)^2}\over \eta} \ln{t_{max}\over t_{cut}},}
where $t_{max}=xs-(\tilde m_1+\tilde m_2)^2$ and $t_{cut}=4 \GeV^2$ are the
limits put on $Q^2$ for the deep-inelastic process. Also, $e_{q_f}$ is the
electric charge of the $q_f$ quark, $x$ is the parton density distribution
variable, and $\eta=z/x$. The total cross section for the deep-inelastic
processes is thus given by \ALTA
\eqn\XII{{\sigma_{deep-inelastic}(e\,parton\to
X\,parton)}=\int^1_{z_{min}}dx\sum_f
q_f(x,Q^2)\int^1_{{1\over x}z_{min}}d\eta
P_{\gamma|q_f}(\eta)\hat \sigma(\hat s),}
where the parton distribution functions of Ref. \MORFIN\ have been used, with
the energy scale $Q^2=(t_{max}-t_{cut})/\ln (t_{max}/t_{cut})$.

It has been observed that by using the WW approximation, the results are
usually larger than the exact results by $20-30\%$ for the elastic case \DREES.
However, for the deep-inelastic processes the WW results are smaller than the
exact ones \ALTA. Consequently the WW approximation will not enhance the
effects and is thus good enough in light of the inherent uncertainties in this
type of calculations. Moreover, these shifts in the cross sections are
equivalent to shifts in the selectron or chargino masses of $5 \GeV$ or less.

\newsec{Results}
\subsec{Selectron-neutralino production}
There are four possible production channels at HERA
\eqn\XIII{ep\to \tilde e_R\chi^0_1,
		   \tilde e_R\chi^0_2,
		   \tilde e_L\chi^0_1,
		   \tilde e_L\chi^0_2+X.}
By far the largest cross section is for the $\tilde e_R\chi^0_1$ channel.
The main reason for this is that in this model the $\tilde e_R$ mass is much
smaller that the $\tilde e_L$ mass (see Eq. \FIII{a}), and the first neutralino
$\chi^0_1$ is the lightest SUSY particle (LSP), which escapes detection. By
the same kinematical reasons the $\tilde e_R \chi^0_2$ and $\tilde e_L\chi^0_1$
cross sections are smaller but still observable, whereas the $\tilde
e_L\chi^0_2$ contribution is negligible ($<10^{-3}\pb$). This pattern holds
for both elastic and deep-inelastic processes. The above sparticles decay
mostly in the following ways
\eqna\XIV
$$\eqalignno{&\tilde e_L\to e_L\chi^0_1,&\XIV a\cr
	    &\tilde e_R\to e_R\chi^0_1,&\XIV b\cr
            &\chi^0_2\to \nu_l\bar\nu_l\chi^0_1, l^+l^-\tilde
\chi^0_1, q\bar q\chi^0_1.&\XIV c\cr}$$
However, in this model there are some points in the parameter space that also
allow the rare decay channels $\tilde e_L\to e_L\chi^0_2$ and $\tilde e_R\to
e_R\chi^0_2$. These only contribute for a small region of parameter space
($\approx 12\%$ of the allowed points) and are phase space suppressed. The
cross section for the dominant elastic $ep\to\tilde e_R\chi^0_1
\to ep+\slp$ and deep-inelastic $ep\to\tilde e_R\chi^0_1\to eX+\slp$ processes
are shown in the top row of Fig. 2 and 3 respectively. Note that for
increasingly larger selectron masses, the cross section for the deep-inelastic
process drops faster than that for the elastic one. The analogous results for
the smaller $\tilde e_R\chi^0_2$ and $\tilde e_L\chi^0_1$ channels are shown in
the bottom row of Figs. 2,3.

Let us consider the four elastic cross sections $\sigma(\tilde
e_{R,L}\chi^0_{1,2})$ in order to disentangle the best signal to be
experimentally detected. According to Ref. \DREES, the cross section for the
elastic processes (Eq. \XIII) peaks at a value ($p^\star_e$) of the daughter
electron transverse momentum given by
\eqn\XVI{p^\star_e={m^2_{\tilde e_{R,L}}-m^2_{\chi^0_{1,2}}
\over 2m_{\tilde e_{R,L}}}.}
Moreover, a Monte-Carlo study shows that the average transverse momentum is
close to $\vev{p^e_T}\approx p^\star_e$. To get an idea of the most likely
values of $p^e_T$, we have computed the average $\tilde p^\star_e$ (weighed by
the four elastic cross sections) and the results are shown in Fig. 4. Clearly,
the daughter electrons will be hard and with large $p_T$. This is an excellent
signal to be detected at HERA.

For elastic processes, another measurable signal is the slowed down outgoing
proton. Since the transverse momentum of the outgoing proton is very small,
the relative energy loss of the proton energy $z=(E^{in}_p-E^{out}_p)/E^{in}_p$
is given by $z=1-x_L$, where $x_L$ is the longitudinal momentum of the leading
proton. It has been pointed out \DREES\ that the $z$-distribution is peaked at
a value not much larger than its minimal value,
\eqn\XVII{z_{min}={1\over s}(m_{\tilde e_{R,L}}+m_{\chi^0_{1,2}})^2.}
Therefore, the smallest measured value in the $z$-distribution should be a good
approximation to $z_{min}$. Since the Leading Proton Spectrometer (LPS) of the
ZEUS detector at HERA can measure this distribution accurately, one may have
a new way of probing the supersymmetric spectrum, as follows. We calculate the
average $\tilde z_{min}$ weighed by the different elastic cross sections
$\sigma(\tilde e_{R,L}\chi^0_{1,2})$. The results are shown in the top row of
Fig. 5 versus the total elastic cross section. These plots show
the possible values of $\tilde z_{min}$ for a given sensitivity. For example,
if elastic cross sections could be measured down to $\approx10^{-3}\pb$, then
$\tilde z_{min}$ could be probed up to $\approx0.2$. Now, $\tilde z_{min}$
can be computed from Eq. \XVII\ and be plotted against, say $m_{\tilde e_R}$,
as shown in the bottom row of Fig. 5. For the example given above ($\tilde
z_{min}\lsim0.2$) one could indirectly probe $\tilde e_R$ masses as high as
$\approx115\GeV$. Note that a useful constraint on $m_{\tilde e_R}$ is possible
because the correlation among the various sparticle masses in this model makes
these scatter plots be rather well defined. This indirect experimental
exploration still requires the identification of elastic supersymmetric events
with $eX+\mpt$ signature (in order to identify protons that contribute to the
relevant $z$-distribution), but does not require a detailed reconstruction of
each such event.

One interesting phenomenon in selectron-neutralino production at HERA is the
possibility of using polarized electron beams. Since we have seen that
$\sigma(\tilde e_R\chi^0_1)\gg\sigma(\tilde e_L\chi^0_{1,2})$, right-handed
beams are expected to be much more active in producing SUSY signals than
left-handed beams. To compare the results obtained with R and L polarized
beams is a further selection power to disentangle a genuine signal at HERA.

\subsec{Sneutrino-chargino production}

Unlike selectron-neutralino production, where right-handedly polarized beam
electrons yield the largest signal, sneutrino-chargino production can only
occur when the electron beam is {\it not} completely right-handedly polarized,
because $\tilde\nu_{e_L}$ couples only to left-handed electrons. The allowed
decay modes for the channel in Eq. \I{b} are
\eqna\XVIII
$$\eqalignno{&\tilde \nu_e\to \chi^0_1\nu_e,
		\chi^0_2\nu_e, \chi^\mp_1 e^\pm_L, &\XVIII a\cr
&\chi^-_1\to \chi^0_1l^-\bar\nu_l, \chi^0_1 q\bar q^\prime.&\XVIII b\cr}$$
Since the masses of $\chi^0_2$ or $\chi^-_1$ are usually larger than the
sneutrino mass, $\tilde \nu_e$ can rarely decay to $\chi^0_2$ or $\chi^-_1$
and thus decays mostly invisibly. To contribute to the desired $eX+\mpt$
signal, the chargino must decay leptonically. In this model this branching
ratio is quite sizeable (see Fig. 2 in Ref. \LNWZ). Moreover, for most points
in the allowed parameter space of the model, the daughter electron from the
decay of the chargino is hard $(E_l>5 \GeV)$. For a detail discussion of this
point, we refer the reader to Ref. \LNWZ. The cross section for this process,
including branching ratios, is shown in Fig. 6 (top (bottom) row for elastic
(deep-inelastic) contribution), and can be seen to be of the same order as that
for selectron-neutralino production ({\it c.f.} Figs. 2,3).

The signature for this production channel is different from the
selectron-neutralino channel in the following ways: (i) it only produces
left-handed daugther leptons (compared to dominantly right-handed ones); and
(ii) the daughter leptons can equally likely be of any flavor (as opposed to
only electrons). Since $\chi^-_1$ can also decay likely into hadronically noisy
jets, in general, sneutrino-chargino detection is more complicated than
selectron-neutralino detection. But ``noise" and ``complications" could be
disentangled since a right-handedly polarized electron
beam would shut off this channel completely.

\newsec{Discussion and conclusion}
We have investigated the relevant SUSY production channels at HERA within
the no-scale flipped $SU(5)$ supergravity model, where direct squark production
is highly suppressed. Because of the different masses of $\tilde e_L$ and
$\tilde e_R$, the production rate is dramatically different when the incident
electron beam is polarized left-handedly or right-handedly. If it is
right-handedly polarized, then the $\tilde e_R\chi^0_{1,2}$ channels will be
the only ones allowed, with a hard electron with large $p_T$ as the dominant
signal. If the beam is left-handedly polarized, only the much smaller $\tilde
e_L\chi^0_{1,2}$ channels will contribute, as well as the hadronically noisy
$\tilde\nu_e\chi^-_1$ channel. This tuning of the machine would be relevant
only after positive sparticle identification. Before that the unpolarized
beam will allow for a larger total supersymmetric signal. In Fig. 7 (top row)
we show the total elastic plus deep-inelatic $\tilde e_{R,L}\chi^0_{1,2}$
signal versus $m_{\tilde e_R}$, which would be relevant for right-handed beam
polarization. In the bottom row of the same figure we show the {\it total}
supersymmetric cross section into $eX+\slp$, including $\tilde
e_{R,L}\chi^0_{1,2}$ and $\tilde\nu_e\chi^-_1$, versus $m_{\tilde e_R}$. This
plot, and its analogs in Fig. 8 where $m_{\tilde e_R}$ is replaced by
$m_{\chi^0_1}$ and $m_{\tilde\nu_e}$, show the discovery potential at HERA
for a given sensitivity.

Assuming optimal experimental efficiencies and a suppressed or subtracted-off
background, with an integrated luminosity of ${\cal L}=100\,(1000)\ipb$, and
demanding at least five fully identified events (\ie,
$\sigma>5\times10^{-2}\,(5\times10^{-3})\pb$), one
could probe as high as $m_{\tilde e_R}\approx70\,(90)\GeV$,
$m_{\chi^0_1}\approx40\,(65)\GeV$, and $m_{\tilde\nu_e}\approx70\,(125)\GeV$.
The analogous plots versus $m_{\chi^\pm_1}$ are not very informative in
pinning down the discovery limit in this variable, since it ranges widely
$m_{\chi^\pm_1}\lsim50-115\,(120-170)\GeV$ for ${\cal L}=100\,(1000)\ipb$.
The short term discovery limits (${\cal L}=100\ipb$) would extend the
present LEPI lower bounds on these sparticle masses by $\approx25\GeV$.
The long term discovery limits are competitive with those foreseable at LEPII
\LNPWZ. We have also shown that the Leading Proton Spectrometer (LPS) at HERA
is an excellent supersymmetry detector which can provide indirect information
about the sparticle masses by measuring the leading proton longitudinal
momentum distribution in elastic $e\slp+p$ processes, without the need to
reconstruct all such events. We conclude that HERA is an interesting
supersymmetric probe in the no-scale flipped $SU(5)$ supergravity model.
\bigskip
\bigskip
\bigskip
\noindent{\it Acknowledgments}: This work has been supported in part by DOE
grant DE-FG05-91-ER-40633. The work of J.L. has been supported by an SSC
Fellowship. The work of  D.V.N. has been supported in part by a grant from
Conoco Inc. The work of X. W. has been supported by a T-1 World-Laboratory
Scholarship. We would like to thank the HARC Supercomputer Center for the use
of their NEC SX-3 supercomputer and the Texas A\&M Supercomputer Center for the
use of their CRAY-YMP supercomputer.
%\listrefsd
\listrefs
%\listfigsd
\listfigs
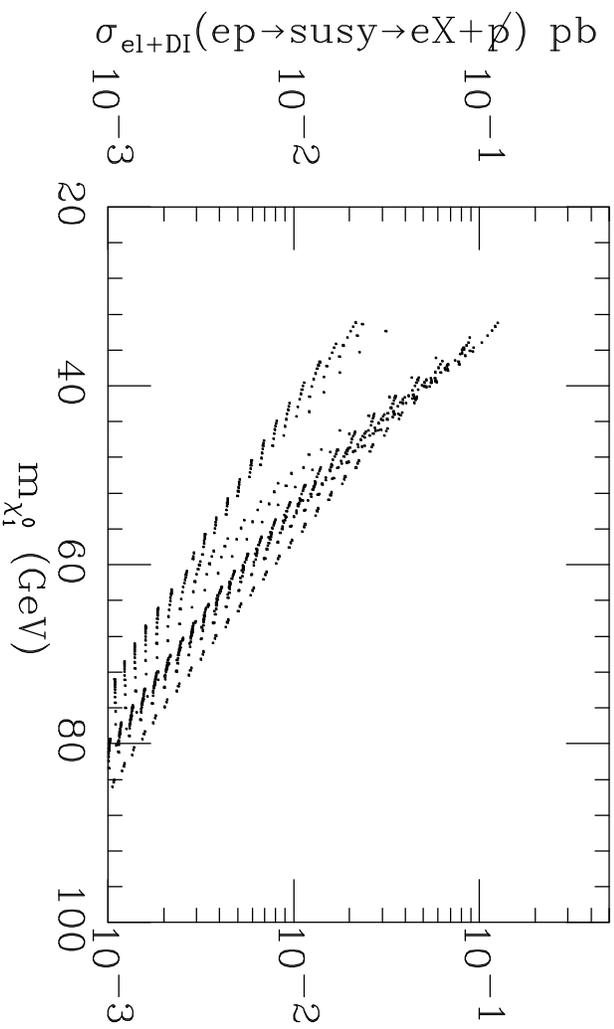
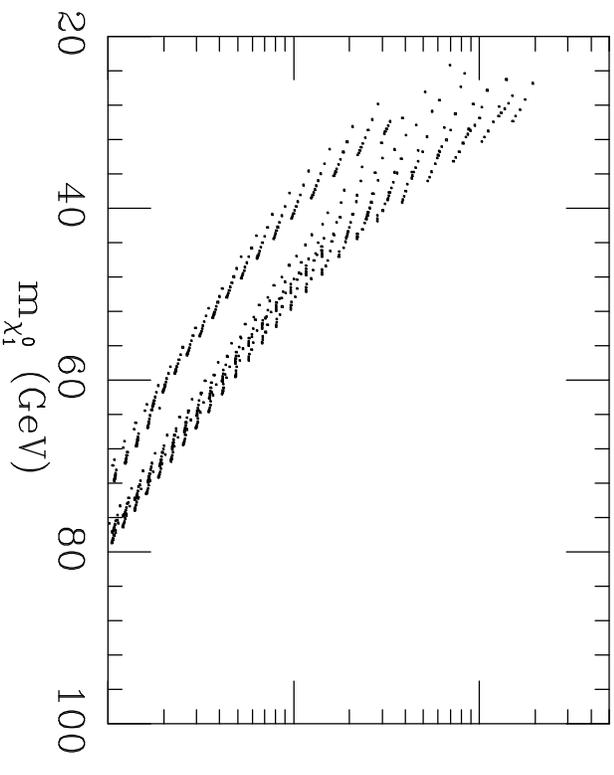
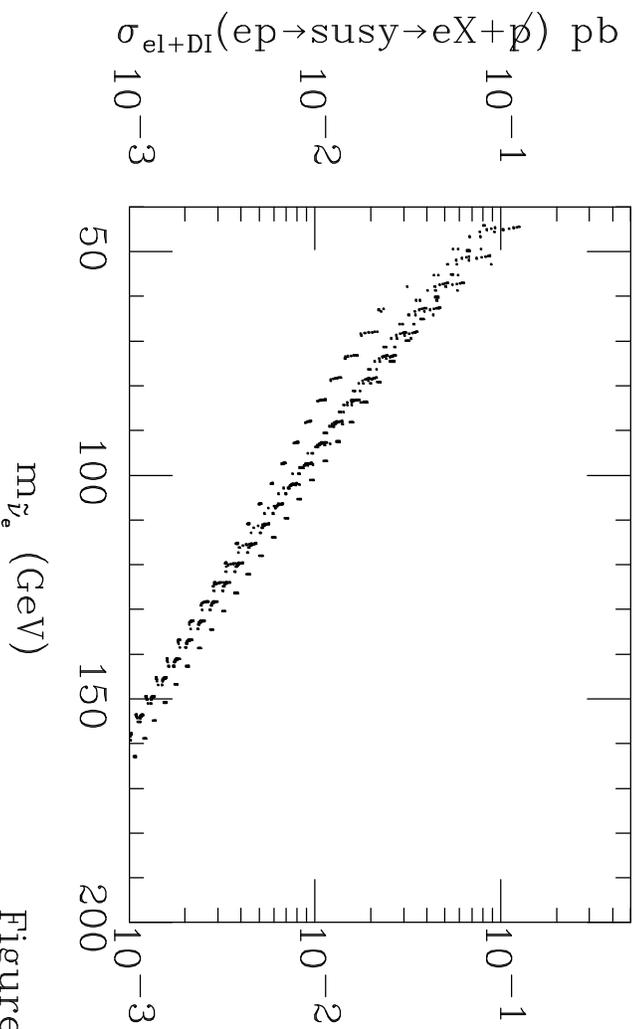
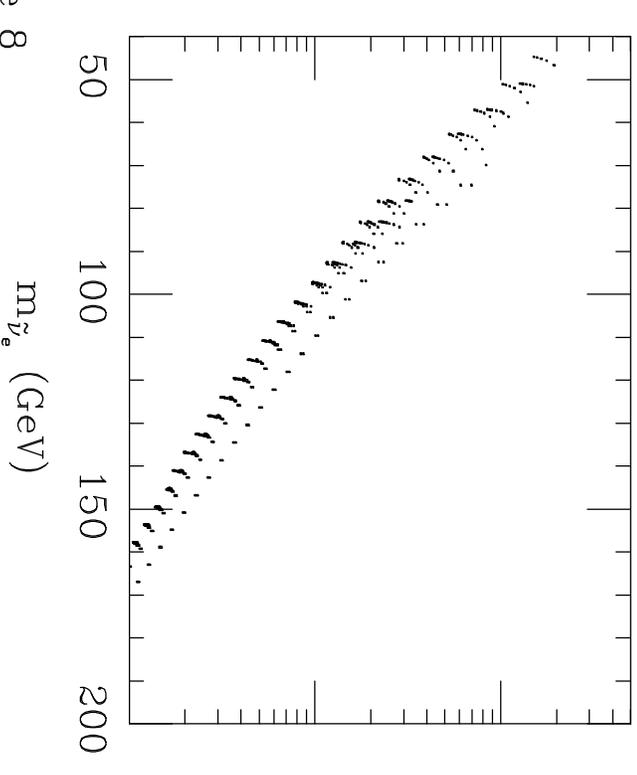
\bye